\documentclass[final,5p,times,twocolumn,nopreprintline]{elsarticle}
\usepackage{amsmath,slashed,booktabs,dsfont}
\usepackage{graphicx,graphics}
\usepackage{dcolumn}
\usepackage[hyperfootnotes=false]{hyperref}
\usepackage{xspace}
\usepackage{color}
\usepackage{balance}
\usepackage{multirow}
\usepackage{multicol}

\usepackage{fancyhdr}
\fancypagestyle{firstpage}{%
	
	\lhead{}
	\rhead{\small PSI-PR-25-10, ZU-TH 27/25}
}
\newcommand{\beq}{\begin{equation}}
\newcommand{\eeq}{\end{equation}}

\newcommand{\Order}{\mathcal{O}}

\newcommand{\GeV}{\,\text{GeV}}

\allowdisplaybreaks[1]

\begin{document}

\renewcommand{\theequation}{\arabic{equation}}

\begin{frontmatter}
 
\title{Hadronic light-by-light scattering in the anomalous magnetic moments of electron and $\tau$}

\author[Bern]{Martin Hoferichter}
\author[Zurich,PSI]{Peter Stoffer}
\author[Bern]{Maximilian Zillinger}

\address[Bern]{Albert Einstein Center for Fundamental Physics, Institute for Theoretical Physics, University of Bern, Sidlerstrasse 5, 3012 Bern, Switzerland}
\address[Zurich]{Physik-Institut, Universit\"at Z\"urich, Winterthurerstrasse 190, 8057 Z\"urich, Switzerland}
\address[PSI]{PSI Center for Neutron and Muon Sciences, 5232 Villigen PSI, Switzerland}

\begin{abstract}
 In Refs.~\cite{Hoferichter:2024vbu,Hoferichter:2024bae} we provided a complete dispersive evaluation of the hadronic light-by-light (HLbL) contribution to the anomalous magnetic moment of the muon. While the evaluation strategy was developed for the kinematic situation determined by the muon mass, a similar approach also applies to the HLbL corrections to the anomalous magnetic moments of the electron and $\tau$ lepton, shifting the sensitivity in the loop integrals to smaller and larger momenta, respectively. In this Letter, we propagate the corresponding uncertainties of the various contributions, obtaining $a_e^\text{HLbL}= 3.51(23)\times 10^{-14}$ and $a_\tau^\text{HLbL}= 3.77(29)\times 10^{-8}$.
\end{abstract}

\end{frontmatter}

\thispagestyle{firstpage}

\section{Introduction}

Motivated by the experimental campaign of the Fermilab experiment, which dominates the current world average for the anomalous magnetic moment of the muon~\cite{Muong-2:2023cdq,Muong-2:2024hpx},
\beq
a_\mu^\text{exp}=116\,592\,059(22)\times 10^{-11},
\eeq
substantial efforts have been invested over the last years to obtain a Standard-Model (SM) prediction at a commensurate level of precision~\cite{Aoyama:2020ynm,Aoyama:2012wk,Aoyama:2019ryr,Czarnecki:2002nt,Gnendiger:2013pva,Davier:2017zfy,Keshavarzi:2018mgv,Colangelo:2018mtw,Hoferichter:2019gzf,Davier:2019can,Keshavarzi:2019abf,Hoid:2020xjs,Kurz:2014wya,Melnikov:2003xd,Colangelo:2014dfa,Colangelo:2014pva,Colangelo:2015ama,Masjuan:2017tvw,Colangelo:2017qdm,Colangelo:2017fiz,Hoferichter:2018dmo,Hoferichter:2018kwz,Gerardin:2019vio,Bijnens:2019ghy,Colangelo:2019lpu,Colangelo:2019uex,Blum:2019ugy,Colangelo:2014qya}, mainly addressing hadronic effects. In the case of hadronic vacuum polarization (HVP), at present, this objective has not yet been met, with ongoing work addressing the tensions among evaluations based on $e^+e^-\to\text{hadrons}$ data~\cite{Davier:2017zfy,Keshavarzi:2018mgv,Colangelo:2018mtw,Hoferichter:2019gzf,Davier:2019can,Keshavarzi:2019abf,Hoid:2020xjs,Crivellin:2020zul,Keshavarzi:2020bfy,Malaescu:2020zuc,Colangelo:2020lcg,Stamen:2022uqh,Colangelo:2022vok,Colangelo:2022prz,Hoferichter:2023sli,Hoferichter:2023bjm,Stoffer:2023gba,Davier:2023fpl,CMD-3:2023alj,CMD-3:2023rfe,Leplumey:2025kvv,Hoferichter:2025lcz} and with lattice QCD~\cite{Borsanyi:2020mff,Ce:2022kxy,ExtendedTwistedMass:2022jpw,FermilabLatticeHPQCD:2023jof,RBC:2023pvn,Boccaletti:2024guq,Blum:2024drk,Djukanovic:2024cmq,Bazavov:2024eou}, e.g., by scrutinizing the
radiative corrections in the critical $e^+e^-\to\pi^+\pi^-$ channel~\cite{Campanario:2019mjh,Ignatov:2022iou,Colangelo:2022lzg,Monnard:2021pvm,Abbiendi:2022liz,BaBar:2023xiy,Aliberti:2024fpq}.

In contrast, for hadronic light-by-light (HLbL) scattering~\cite{Melnikov:2003xd,Masjuan:2017tvw,Colangelo:2017qdm,Colangelo:2017fiz,Hoferichter:2018dmo,Hoferichter:2018kwz,Gerardin:2019vio,Bijnens:2019ghy,Colangelo:2019lpu,Colangelo:2019uex,Pauk:2014rta,Danilkin:2016hnh,Jegerlehner:2017gek,Knecht:2018sci,Eichmann:2019bqf,Roig:2019reh} the uncertainty compared to Ref.~\cite{Aoyama:2020ynm} could be substantially reduced, on the one hand, thanks to more precise lattice-QCD calculations~\cite{Blum:2019ugy,Chao:2021tvp,Chao:2022xzg,Blum:2023vlm,Fodor:2024jyn} and improved phenomenological evaluations on the other. In particular, further exclusive hadronic states were evaluated~\cite{Hoferichter:2020lap,Zanke:2021wiq,Danilkin:2021icn,Stamen:2022uqh,Ludtke:2023hvz,Hoferichter:2023tgp,Hoferichter:2024fsj,Ludtke:2024ase,Deineka:2024mzt,Holz:2024lom,Holz:2024diw} in a dispersive approach to HLbL scattering~\cite{Colangelo:2014dfa,Colangelo:2014pva,Colangelo:2015ama,Hoferichter:2013ama}, higher-order short-distance constraints derived~\cite{Bijnens:2020xnl,Bijnens:2021jqo,Bijnens:2022itw,Bijnens:2024jgh}, and their matching onto hadronic descriptions improved~\cite{Leutgeb:2019gbz,Cappiello:2019hwh,Knecht:2020xyr,Masjuan:2020jsf,Ludtke:2020moa,Colangelo:2021nkr,Leutgeb:2021mpu,Colangelo:2024xfh,Leutgeb:2024rfs,Mager:2025pvz}. Profiting from these developments, we presented a complete dispersive evaluation of the HLbL contribution to $a_\mu$ in Refs.~\cite{Hoferichter:2024vbu,Hoferichter:2024bae}
\beq
a_\mu^\text{HLbL}=101.9(7.9)\times 10^{-11}.
\eeq
The evaluation strategy was optimized for the kinematic configuration set by the muon mass, but, in principle, within a dispersive approach a similar calculation is possible for arbitrary lepton mass, once the HLbL tensor is determined. The key question becomes how the error propagation changes when the lepton mass is varied, given the large hierarchies of $m_\mu/m_e=206.7682830(46)$ and $m_\tau/m_\mu=16.8170(11)$~\cite{Tiesinga:2021myr}. Accordingly, much lower/higher loop momenta will dominate, and thus shift the focus to other momentum regions of the HLbL tensor. In this Letter, we study this error propagation using the same formalism developed for $a_\mu^\text{HLbL}$, to obtain improved values of $a_e^\text{HLbL}$ and $a_\tau^\text{HLbL}$.

For $a_e$, the overall uncertainty is not actually dominated by the hadronic effects---HVP, HLbL, and higher-order iterations~\cite{Calmet:1976kd,Kurz:2014wya,Colangelo:2014qya,Hoferichter:2021wyj}---but by independent input for the fine-structure constant $\alpha$, for which determinations from Cs~\cite{Parker:2018vye}  and Rb~\cite{Morel:2020dww} atom interferometry differ by $5.4\sigma$, limiting the beyond-the-SM sensitivity of the experimental result~\cite{Fan:2022eto}
\beq
\label{ae_exp}
a_e^\text{exp}=115\,965\,218\,059(13)\times 10^{-14}
\eeq
due to $a_e^\text{SM}[\text{Cs}]-a_e^\text{SM}[\text{Rb}]\simeq 1.36(25)\times 10^{-12}$.
Meanwhile, since a tension in the five-loop QED coefficient has been resolved recently~\cite{Aoyama:2019ryr,Volkov:2019phy,Volkov:2024yzc,Aoyama:2024aly}, the main theory uncertainty again arises from HVP, at a level of $\simeq 3\times 10^{-14}$~\cite{DiLuzio:2024sps}. Estimates for the HLbL correction were given previously~\cite{Jegerlehner:2017gek,Jegerlehner:2009ry,Prades:2009tw}:
\begin{align}
\label{ae_HLbL_Jegerlehner}
a_e^\text{HLbL}\text{\cite{Jegerlehner:2017gek}}&=3.7(5)\times 10^{-14},\notag\\
a_e^\text{HLbL}\text{\cite{Jegerlehner:2009ry}}&=3.9(1.3)\times 10^{-14},\notag\\
a_e^\text{HLbL}\text{\cite{Prades:2009tw}}&=3.5(1.0)\times 10^{-14},
\end{align}
only a factor $3.5$ in size below the uncertainty in Eq.~\eqref{ae_exp}. It is therefore interesting to check if Eq.~\eqref{ae_HLbL_Jegerlehner} is consistent with a complete evaluation.

The SM prediction for $a_\tau$~\cite{Keshavarzi:2019abf} also includes an estimate of the HLbL correction~\cite{Eidelman:2007sb,Eidelman:2016aih}
\beq
\label{atau_HLbL_Passera}
a_\tau^\text{HLbL}=5(3)\times 10^{-8},
\eeq
which even amounts to the largest source of uncertainty. Accordingly, an improved evaluation would have a direct impact on the precision of the SM prediction. Unfortunately, despite recent measurements in peripheral Pb--Pb collisions at LHC~\cite{ATLAS:2022ryk,CMS:2022arf,CMS:2024qjo}, the precision of $a_\tau^\text{exp}$ does not even allow for a test of leading-order QED, rendering a sensitivity below $10^{-7}$ out of reach for the foreseeable future. However, given recent developments aiming at extracting substantially improved constraints on $a_\tau$ from polarized $e^+e^-\to\tau^+\tau^-$ scattering at Belle~II in case of a potential SuperKEKB upgrade~\cite{Bernabeu:2007rr,Crivellin:2021spu,USBelleIIGroup:2022qro,Aihara:2024zds,Gogniat:2025eom}, it is still worthwhile to complete its SM prediction with a refined calculation of  $a_\tau^\text{HLbL}$.

As reference point, we first review the salient features of the dispersive calculation from Refs.~\cite{Hoferichter:2024vbu,Hoferichter:2024bae} in Sec.~\ref{sec:disp}, which we then subsequently adapt to $\ell=e,\tau$ in Sec.~\ref{sec:uds}. For the charm loop, see Sec.~\ref{sec:charm}, we present a slightly improved evaluation compared to the estimate from Ref.~\cite{Colangelo:2019uex}, following the strategy from Ref.~\cite{Hoferichter:2025yih}. We conclude in Sec.~\ref{sec:conclusions}.

\begin{table}[t]
	\centering
	\renewcommand{\arraystretch}{1.3}
	\scalebox{0.95}{\begin{tabular}{lrrr}
	\toprule
	 Contribution & $a_e\ [10^{-14}]$ & $a_{\tau}\ [10^{-8}]$ & Reference\\\midrule
     $\pi^0$ & $2.72^{+0.09}_{-0.07}$ & $0.65^{+0.06}_{-0.04}$ & \cite{Hoferichter:2018dmo,Hoferichter:2018kwz}\\
     $\eta$ & $0.50(3)$ & $0.23(2)$ & \cite{Holz:2024lom,Holz:2024diw}\\
     $\eta'$ & $0.41(2)$ & $0.30(3)$ & \cite{Holz:2024lom,Holz:2024diw}\\
	 $\pi^0$, $\eta$, $\eta'$ poles & $3.63^{+0.10}_{-0.08}$ & $1.19^{+0.07}_{-0.05}$ &\cite{Hoferichter:2018dmo,Hoferichter:2018kwz,Holz:2024lom,Holz:2024diw}\\
	 $\pi^\pm$ box & $-0.66(1)$ & $-0.13(1)$ & \cite{Colangelo:2017qdm,Colangelo:2017fiz}\\
	 $K^\pm$ box & $-0.02(0)$ & $-0.01(0)$ &\cite{Stamen:2022uqh}\\
	 $S$-wave rescattering & $-0.32(5)$ & $-0.11(1)$ &\cite{Colangelo:2017qdm,Colangelo:2017fiz,Danilkin:2021icn,Deineka:2024mzt}\\\midrule
	 Sum & $2.63^{+0.11}_{-0.10}$ & $0.94^{+0.07}_{-0.05}$ &\\
\bottomrule
	\renewcommand{\arraystretch}{1.0}
	\end{tabular}}
	\caption{Summary of the lowest hadronic states contributing to $a_\ell^\text{HLbL}$, integrated over the entire domain.}
	\label{tab:disp_summary}
\end{table}

\section{Dispersive formalism}
\label{sec:disp}

The calculation is based on a dispersive approach to HLbL scattering~\cite{Colangelo:2014dfa,Colangelo:2014pva,Colangelo:2015ama,Colangelo:2017fiz,Hoferichter:2013ama}, in which the HLbL tensor $\Pi^{\mu\nu\lambda\sigma}$ is reconstructed from its singularities, combined with short-distance constraints (SDCs) for the asymptotic behavior and mixed momentum regions in the loop integral. First, the HLbL tensor is reduced to scalar function $\Pi_i$ via a Bardeen--Tung--Tarrach~\cite{Bardeen:1968ebo,Tarrach:1975tu} Lorentz decomposition, leading to the master formula
\begin{align}	\label{eq:MasterFormulaPolarCoord}
	a_\mu^\text{HLbL} &= \frac{\alpha^3}{432\pi^2} \int_0^\infty d\Sigma\, \Sigma^3 \int_0^1 dr\, r\sqrt{1-r^2} \int_0^{2\pi} d\phi \notag\\
	&\qquad \times\sum_{i=1}^{12} T_i(\Sigma,r,\phi) \bar\Pi_i(q_1^2,q_2^2,q_3^2),
\end{align}
with kernel functions $T_i(\Sigma,r,\phi)$~\cite{Colangelo:2017fiz}, $\bar\Pi_i$ referring to a subset of the $\Pi_i$, and photon virtualities $Q_i^2=-q_i^2$~\cite{Eichmann:2015nra}
\begin{align}
\label{Qi}
		Q_{1/2}^2 &= \frac{\Sigma}{3} \left( 1 - \frac{r}{2} \cos\phi \mp \frac{r}{2}\sqrt{3} \sin\phi \right), \notag\\
		Q_3^2 &= \frac{\Sigma}{3} \left( 1 + r \cos\phi \right).
\end{align}
For the calculation of the $\bar\Pi_i$ we follow Refs.~\cite{Hoferichter:2024vbu,Hoferichter:2024bae}. For low virtualities $Q_i\leq Q_0$, below a matching scale $Q_0$, we use exclusive hadronic states, i.e., (i) $\pi^0$, $\eta$, $\eta'$ poles, determined by their transition form factors (TFFs)~\cite{Stollenwerk:2011zz,Schneider:2012ez,Hoferichter:2012pm,Hanhart:2013vba,Hoferichter:2014vra,Kubis:2015sga,Holz:2015tcg,Hoferichter:2018dmo,Hoferichter:2018kwz,Hoferichter:2021lct,Holz:2022hwz,Holz:2022smu,Holz:2024lom,Holz:2024diw}; (ii) two-meson intermediate states, including so-called box and rescattering contributions~\cite{Colangelo:2017qdm,Colangelo:2017fiz,Danilkin:2021icn,Stamen:2022uqh,Deineka:2024mzt}, as determined, respectively, by the corresponding electromagnetic form factors and partial waves for $\gamma^*\gamma^*\to\pi\pi/\bar K K/\pi \eta$~\cite{Garcia-Martin:2010kyn,Hoferichter:2011wk,Moussallam:2013una,Hoferichter:2013ama,Danilkin:2018qfn,Hoferichter:2019nlq,Danilkin:2019opj,Lu:2020qeo,Schafer:2023qtl,Deineka:2024mzt}; (iii) axial-vector resonances, with TFFs from Refs.~\cite{Zanke:2021wiq,Hoferichter:2023tgp}, constrained by data for $e^+e^-\to e^+e^- A$, $A=f_1,f_1'$~\cite{Achard:2001uu,Achard:2007hm}, $e^+e^-\to f_1\pi^+\pi^-$~\cite{BaBar:2007qju,BaBar:2022ahi}, and radiative decays~\cite{Zanke:2021wiq,Hoferichter:2023tgp,ParticleDataGroup:2024cfk} as well as U(3) symmetry; (iv) heavy scalar and tensor states estimated with quark-model-inspired TFFs~\cite{Schuler:1997yw,Hoferichter:2020lap}.

\begin{table}[t]
	\centering
	\renewcommand{\arraystretch}{1.3}
	\scalebox{1.00}{\begin{tabular}{lrrrrrr}
	\toprule
	$[10^{-14}]$ & $a_e[\bar\Pi_{1,2}] $ & $a_e[\bar\Pi_{3\text{--}12}] $ & Sum \\\midrule
LO & $0.14$ & $0.04$ & $0.18$ \\
NLO $\times\frac{\pi}{\alpha_s(Q_0)}$ & $-0.13$ & $-0.05$ & $-0.18$ \\
LO+NLO & $0.13^{+0.00}_{-0.01}$ & $0.04^{+0.00}_{-0.01}$ & $0.16^{+0.00}_{-0.01}$ \\\midrule
$[10^{-8}]$ &  $a_\tau[\bar\Pi_{1,2}]$ & $a_\tau[\bar\Pi_{3\text{--}12}]$ & Sum \\\midrule
LO  & $0.93$ & $0.34$ & $1.26$\\
NLO $\times\frac{\pi}{\alpha_s(Q_0)}$  & $-0.86$ & $-0.38$ & $-1.25$\\
LO+NLO  & $0.83^{+0.01}_{-0.03}$ & $0.30^{+0.01}_{-0.01}$ & $1.12^{+0.03}_{-0.04}$\\
\bottomrule
	\renewcommand{\arraystretch}{1.0}
	\end{tabular}}
	\caption{Quark loop and $\alpha_s$ corrections to $a_\ell^\text{HLbL}$ for $Q_0=1.5\GeV$~\cite{Bijnens:2021jqo}, separately for the scalar functions $\bar\Pi_{1,2}$, $\bar\Pi_{3\text{--}12}$, and their sum. The scale in $\alpha_s(\mu)$ is varied within $[Q_0/\sqrt{2},Q_0\sqrt{2}]$, the decoupling scales $\mu_c=3\GeV$, $\mu_b=m_b$ by a factor $2$, leading to $\alpha_s(Q_0)=0.35^{+0.11}_{-0.06}$~\cite{Herren:2017osy,Chetyrkin:2000yt}.}
	\label{tab:quark_loop}
\end{table}

\begin{table*}[t!]
	\centering
	\scalebox{0.836}{
	\renewcommand{\arraystretch}{1.3}
	\begin{tabular}{llrrrrrr}
	\toprule
	 Region & & $a_e[\bar\Pi_{1,2}] \ [10^{-14}]$ & $a_e[\bar\Pi_{3\text{--}12}] \ [10^{-14}]$ & Sum $[10^{-14}]$ & $a_{\tau}[\bar\Pi_{1,2}] \ [10^{-8}]$ & $a_{\tau}[\bar\Pi_{3\text{--}12}] \ [10^{-8}]$ & Sum $[10^{-8}]$\\\midrule
\multirow{8}{*}{$Q_i<Q_0$} 
& $P=\pi^0,\eta,\eta'$ &  $3.41(8)_\text{low}$ & -- & $3.41(8)_\text{low}$ & $0.77(3)_\text{low}$ & -- & $0.77(3)_\text{low}$\\
& $\pi^\pm$ box &  $0.23(0)_\text{low}$ & $-0.88(1)_\text{low}$ & $-0.66(1)_\text{low}$ & $0.03(0)_\text{low}$ & $-0.16(1)_\text{low}$ & $-0.13(1)_\text{low}$\\
& $K^\pm$ box &  $0.01(0)_\text{low}$ & $-0.02(0)_\text{low}$ & $-0.01(0)_\text{low}$ & $0.00(0)_\text{low}$ & $-0.01(0)_\text{low}$ & $-0.01(0)_\text{low}$\\
& $S$-wave rescattering & -- & $-0.31(5)_\text{low}$ & $-0.31(5)_\text{low}$ & -- & $-0.10(1)_\text{low}$ & $-0.10(1)_\text{low}$ \\
& $A=f_1,f_1',a_1$ &  $0.19(4)_\text{exp}$ & $0.13(3)_\text{exp}$ & $0.32(6)_\text{exp}$ & $0.16(4)_{\text{exp}}$ & $0.15(3)_{\text{exp}}$ & $0.31(6)_{\text{exp}}$\\
& $S=f_0(1370),a_0(1450)$ & -- & $-0.02(1)_\text{exp}$ & $-0.02(1)_\text{exp}$ & -- & $-0.01(0)_\text{exp}$ & $-0.01(0)_\text{exp}$ \\
& $T=f_2,a_2,f_2'$ & $0.09(1)_\text{exp}$ & $-0.18(2)_\text{exp}$ & $-0.09(1)_\text{exp}$ & $0.03(0)_\text{exp}$ & $-0.06(1)_\text{exp}$ & $-0.03(0)_\text{exp}$\\
& Effective poles & $0.07$ & $-0.01$ & $0.06$ & $0.06$ & $-0.01$ & $0.05$\\\midrule
\multirow{3}{*}{Side $Q_1$} 
& $P=\pi^0,\eta,\eta'$ &  $0.01(0)_\text{low}$ & -- & $0.01(0)_\text{low}$ & $0.02(0)_\text{low}$ & -- & $0.02(0)_\text{low}$\\
& $A=f_1,f_1',a_1$ & $0.01(0)_\text{exp}$ & $0.01(0)_\text{exp}$ & $0.02(0)_\text{exp}$ & $0.02(1)_\text{exp}$ & $0.02(1)_\text{exp}$ & $0.04(1)_\text{exp}$\\
&Effective poles & $0.00$ & $0.00$& $0.00$ & $0.01$ & $0.00$& $0.01$\\\midrule
\multirow{3}{*}{Side $Q_2$} 
& $P=\pi^0,\eta,\eta'$ &  $0.00(0)_\text{low}$ & -- & $0.00(0)_\text{low}$ &  $0.01(0)_\text{low}$ & -- & $0.01(0)_\text{low}$\\
& $A=f_1,f_1',a_1$ & $0.01(0)_\text{exp}$ & $0.00(0)_\text{exp}$ & $0.01(0)_\text{exp}$ & $0.01(0)_\text{exp}$ & $0.01(0)_\text{exp}$ & $0.02(0)_\text{exp}$\\
&Effective poles & $0.00$ & $0.00$& $0.00$ & $0.01$ & $0.00$& $0.01$\\\midrule
\multirow{3}{*}{Side $Q_3$} 
& $P=\pi^0,\eta,\eta'$ &  $0.02(0)_\text{low}$ & -- & $0.02(0)_\text{low}$ &  $0.04(0)_\text{low}$ & -- & $0.04(0)_\text{low}$\\
& $A=f_1,f_1',a_1$ & $0.02(0)_\text{exp}$ & $0.02(0)_\text{exp}$ & $0.04(1)_\text{exp}$ & $0.04(1)_\text{exp}$ & $0.05(1)_\text{exp}$ & $0.09(2)_\text{exp}$\\
&Effective poles & $0.01$ & $0.00$& $0.01$ & $0.02$ & $0.00$& $0.02$\\\midrule
\multirow{4}{*}{Corner $Q_1$} 
& $P=\pi^0,\eta,\eta'$ &  $0.00(0)_\text{low}$ & -- & $0.00(0)_\text{low}$ &  $0.01(0)_\text{low}$ & -- & $0.01(0)_\text{low}$\\
& $A=f_1,f_1',a_1$ & $0.01(0)_\text{exp}$ & $0.00(0)_\text{exp}$ & $0.01(0)_\text{exp}$ & $0.02(1)_\text{exp}$ & $0.01(0)_\text{exp}$ & $0.04(1)_\text{exp}$\\
&Effective poles & $0.01$ & $0.00$& $0.01$ & $0.02$ & $0.00$& $0.02$\\
&OPE & $0.00$ & $0.08$& $0.08$ & $0.00$ & $0.24$& $0.24$\\\midrule
\multirow{4}{*}{Corner $Q_2$} 
& $P=\pi^0,\eta,\eta'$ &  $0.01(0)_\text{low}$ & -- & $0.01(0)_\text{low}$ &  $0.03(0)_\text{low}$ & -- & $0.03(0)_\text{low}$\\
& $A=f_1,f_1',a_1$ & $0.01(0)_\text{exp}$ & $0.00(0)_\text{exp}$ & $0.02(0)_\text{exp}$ & $0.04(1)_\text{exp}$ & $0.01(0)_\text{exp}$ & $0.05(1)_\text{exp}$\\
&Effective poles & $0.01$ & $0.00$& $0.01$ & $0.02$ & $0.00$& $0.02$\\
&OPE & $0.31$ & $0.03$& $0.35$ & $0.54$ & $0.13$& $0.67$\\\midrule
\multirow{4}{*}{Corner $Q_3$} 
& $P=\pi^0,\eta,\eta'$ &  $0.00(0)_\text{low}$ & -- & $0.00(0)_\text{low}$ &  $0.01(0)_\text{low}$ & -- & $0.01(0)_\text{low}$\\
& $A=f_1,f_1',a_1$ & $0.00(0)_\text{exp}$ & $0.00(0)_\text{exp}$ & $0.00(0)_\text{exp}$ & $0.01(0)_\text{exp}$ & $0.00(0)_\text{exp}$ & $0.02(0)_\text{exp}$\\
&Effective poles & $0.00$ & $0.00$& $0.00$ & $0.01$ & $0.00$& $0.01$\\
&OPE & $0.00$ & $0.00$& $0.00$ & $0.01$ & $0.01$& $0.01$\\\midrule
$Q_i>Q_0$ & pQCD & $0.13^{+0.00}_{-0.01}$ & $0.04^{+0.00}_{-0.01}$ & $0.16^{+0.00}_{-0.01}$ & $0.83^{+0.01}_{-0.03}$ & $0.30^{+0.01}_{-0.01}$ & $1.12^{+0.03}_{-0.04}$\\\midrule
\multirow{2}{*}{Sum} & & $4.56(8)_\text{low}(5)_\text{exp}$ & $-1.11(5)_\text{low}(4)_\text{exp}$ & $3.45(9)_\text{low}(8)_\text{exp}$ & $2.78(3)_\text{low}(9)_\text{exp}$ & $0.58(1)_\text{low}(5)_\text{exp}$ & $3.36(4)_\text{low}(13)_\text{exp}$\\
& & $(8)_\text{sys}(9)_\text{eff}$ & $(7)_\text{sys}(7)_\text{eff}$ & $(15)_\text{sys}(11)_\text{eff}$ & $(9)_\text{sys}(10)_\text{eff}$ & $(8)_\text{sys}(2)_\text{eff}$ & $(17)_\text{sys}(11)_\text{eff}$\\
\bottomrule
	\renewcommand{\arraystretch}{1.0}
	\end{tabular}}
	\caption{Summary of the various contributions in the different kinematic regions, at the matching scale $Q_0=1.5\GeV$ and with the OPE applied for $Q_3^2<r(Q_1^2+Q_2^2)/2$, $r={1/4}$, and crossed, using the master formula~\eqref{eq:MasterFormulaPolarCoord}. If not listed, the respective contribution is negligible, e.g., several exclusive states are only relevant for $Q_i<Q_0$. The total uncertainty is obtained by adding to the error components listed the matching error from the variation in $Q_0$ and $r$ (all in quadrature). The mixed region has been divided into $3$ side and $3$ corner regions according to Ref.~\cite{Bijnens:2021jqo}, where, e.g., ``Side $Q_{1}$'' corresponds to $Q_{1}\geq Q_{0},\,Q_{2,3}\leq Q_{0}$ and, e.g., ``Corner $Q_{1}$'' is defined as $Q_{1}\leq Q_{0},\,Q_{2,3}\geq Q_{0}$. The qualitative observations from this further decomposition of the mixed regions also apply to the case of the muon, e.g., for the OPE contribution the numbers from Table 4 in Ref.~\cite{Hoferichter:2024bae} decompose as $a_\mu^\text{OPE}[\bar\Pi_{1,2}]=(-0.2+6.5+0.0)\times 10^{-11}$, $a_\mu^\text{OPE}[\bar\Pi_{3\text{--}12}]=(3.4+1.3+0.0)\times 10^{-11}$, $a_\mu^\text{OPE}[\bar\Pi_{1\text{--}12}]=(3.1+7.8+0.0)\times 10^{-11}$ onto Corner $Q_1+Q_2+Q_3$ (note that in this case the contributions from the pseudoscalar poles are subtracted).}
	\label{tab:subleading}
\end{table*}

Beyond some scale $Q_0$, these hadronic descriptions need to be matched onto SDCs, see, e.g., Refs.~\cite{Leutgeb:2019gbz,Cappiello:2019hwh,Knecht:2020xyr,Masjuan:2020jsf,Ludtke:2020moa,Colangelo:2021nkr,Leutgeb:2024rfs,Eichmann:2024glq}, we again follow Refs.~\cite{Hoferichter:2024vbu,Hoferichter:2024bae}. In the region $Q_i\geq Q_0$, we use perturbative QCD (pQCD) including $\alpha_s$ corrections~\cite{Bijnens:2021jqo} (and the $\alpha_s$ implementation from Refs.~\cite{Herren:2017osy,Chetyrkin:2000yt}). In parts of the mixed regions, such as $Q_3^2\ll (Q_1^2+Q_2^2)/2$, the 
operator product expansion (OPE) relates the HLbL tensor to the 
$VVA$ correlator~\cite{Melnikov:2003xd,Vainshtein:2002nv,Knecht:2003xy}
, a relation that becomes particularly powerful due to a cancellation of certain subleading corrections~\cite{Bijnens:2024jgh}. We use the dispersive calculation for the triplet OPE form factors from Ref.~\cite{Ludtke:2024ase} as well as its generalization to octet and singlet from Refs.~\cite{Hoferichter:2024vbu,Hoferichter:2024bae}. The OPE result is employed when
\beq
Q_3^2\leq r \frac{Q_1^2+Q_2^2}{2},\quad Q_1^2\geq Q_0^2,\quad Q_2^2\geq Q_0^2,\quad Q_{3}^{2}\leq Q_{0}^{2},
\eeq
and crossed versions thereof. The parameters are varied within $r\in[1/8,1/2]$, $Q_0\in[1.2,2.0]\GeV$, and central values quoted for $r=1/4$, $Q_0=1.5\GeV$. Finally, to capture potential low-energy effects propagated from a tower of hadronic states necessary to realize the SDCs, we add a set of effective poles, whose coefficients are matched to the SDCs either in symmetric or asymmetric asymptotic kinematics, while the scales in the TFFs are varied in the same range as $Q_0$. Accordingly, the final error is split into components reflecting uncertainties directly propagated from the experimental input and pQCD (``exp''), the sensitivity to variations of $Q_0$, $r$ (``match''), systematic effects due to the use of U(3) assumptions and simplified tensor TFFs (``sys''), uncertainties due to the effective poles (``eff'').  Moreover, we will show separately the uncertainty derived from the lowest hadronic states as summarized in Table~\ref{tab:disp_summary}, i.e.,  pseudoscalar poles, pion- and kaon-box, and $S$-wave rescattering (``low'').

\section{Light-quark loops}
\label{sec:uds}

In principle, with the HLbL tensor constructed as described in Sec.~\ref{sec:disp}, it is then straightforward to evaluate the loop integral for a different lepton mass and propagate the uncertainties accordingly, see Table~\ref{tab:disp_summary} for the leading hadronic contributions and Table~\ref{tab:quark_loop} for the pQCD quark loop. In practice, however, the numerical techniques need to be adapted. 

\begin{figure*}[t]
     \centering
     \includegraphics[width=0.49\linewidth]{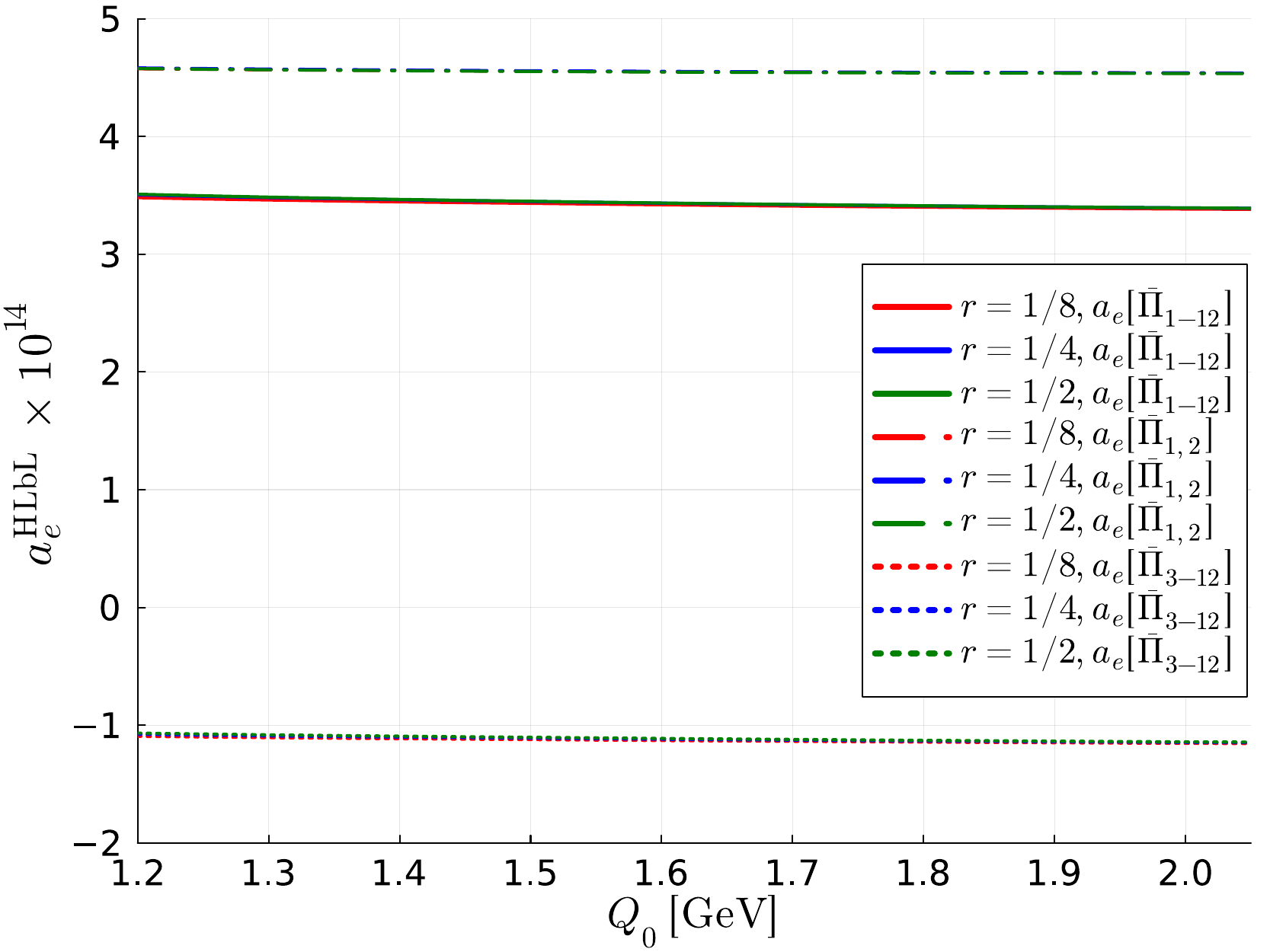}
     \includegraphics[width=0.49\linewidth]{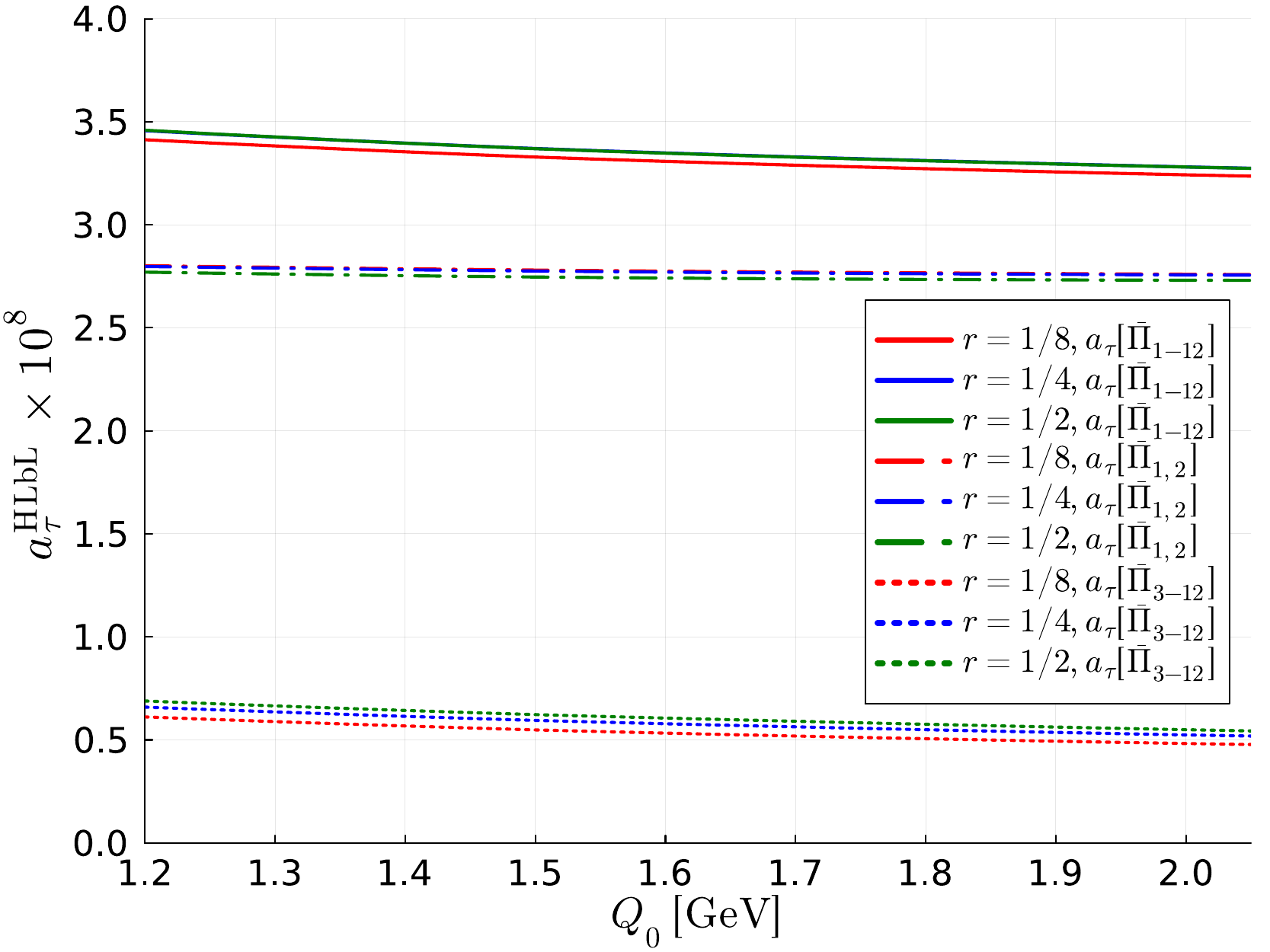}
     \caption{Stability of $a_e^\text{HLbL}$ (left) and $a_\tau^\text{HLbL}$ (right) under variation of $Q_0$, $r$.}
     \label{fig:Q0r}
 \end{figure*}

That is, in the kernel functions $T_{i}(Q_1,Q_2,\tau,m_{\ell})$ 
the muon mass needs to be replaced, which for the $\tau$ lepton does not induce any complications, given that $m_\tau=\Order(1\GeV)$, similarly to other scales in the problem. In contrast, for the electron, numerical instabilities arise, which can be traced back to expressions such as 
\beq
\frac{1}{m_{\ell}^{2}}(1-\sigma_{i})=\frac{1}{m_{\ell}^{2}}\Bigg(1-\sqrt{1+\frac{4m_{\ell}^{2}}{Q_{i}^{2}}}\Bigg)
\eeq
or
\beq
    z = \frac{Q_{1}Q_{2}}{4m_{\ell}^{2}}\big(1-\sigma_{1}\big)\big(1-\sigma_{2}\big),
\eeq
whose evaluation becomes unstable for $m_\ell\to 0$ due to the large cancellations involved. 
In order to avoid or resolve these cancellations numerically, one can either expand the problematic expressions within the kernel functions or use a higher internal precision for their computation. We implemented both approaches, finding numerical agreement. For the latter approach, we were using the julia programming language with ``DoubleFloats'' corresponding to roughly $31$ decimal digits of precision~\cite{Sarnoff_DoubleFloats_2022}. Evaluated at this higher precision, we again observe a very smooth saturation with the cutoff in the master formula~\eqref{eq:MasterFormulaPolarCoord}. Moreover, we verified our charm-loop calculation against the analytic result for heavy-fermion loops~\cite{Kuhn:2003pu}. Empirically, we observe that numerical instabilities only affect the pQCD, OPE, and the charm-loop contributions. The full results are collected in Table~\ref{tab:subleading}. 

For the electron the impact of the pseudoscalar poles is even more enhanced and is by far the most dominant contribution, as it should be, since the low-energy regime is enhanced due to the small electron mass. Moreover, the tensor contribution, which constitutes the most uncertain correction of all narrow resonances, is relatively enhanced compared to the axial-vector contribution. As a result, the associated uncertainty increases accordingly. The mixed and asymptotic regimes are relatively reduced in size. 

For the $\tau$, the biggest change stems from the enhancement of the OPE and pQCD contribution, both being more relevant than the pseudoscalar poles $\pi^{0}$, $\eta$, $\eta'$ combined, which is a direct consequence of the new scale $m_\tau$ in the loop integral. Compared to the situation with the electron or muon, the tensor contribution is now of minor size and hence the corresponding uncertainty of minor importance. The biggest source of uncertainty still comes from the systematic error followed by the uncertainty propagated from the experimental input.  

Finally, the stability of the results with respect to the variation of $Q_0$ and $r$ is illustrated in Fig.~\ref{fig:Q0r}, leading to the total light-quark contributions for the electron (in units of $10^{-14}$)
 \begin{align}
\label{result_uds_e}
 a_e[\bar\Pi_{1,2}]&=4.56(8)_\text{low}(5)_\text{exp}(2)_\text{match}(8)_\text{sys}(9)_\text{eff}[16]_\text{total},\\
 a_e[\bar\Pi_{3\text{--}12}]&=-1.11(5)_\text{low}(4)_\text{exp}(4)_\text{match}(7)_\text{sys}(7)_\text{eff}[13]_\text{total},\notag\\
 a_e[\bar\Pi_{1\text{--}12}]&=3.45(9)_\text{low}(8)_\text{exp}(6)_\text{match}(15)_\text{sys}(11)_\text{eff}[23]_\text{total},\notag
 \end{align}
 and for the $\tau$ (in units of $10^{-8}$)
 \begin{align}
 \label{result_uds_tau}
 a_\tau[\bar\Pi_{1,2}]&=2.78(3)_\text{low}(9)_\text{exp}(5)_\text{match}(10)_\text{sys}(10)_\text{eff}[18]_\text{total},\\
 a_\tau[\bar\Pi_{3\text{--}12}]&=0.58(1)_\text{low}(5)_\text{exp}(11)_\text{match}(8)_\text{sys}(2)_\text{eff}[15]_\text{total},\notag\\
 a_\tau[\bar\Pi_{1\text{--}12}]&=3.36(4)_\text{low}(13)_\text{exp}(13)_\text{match}(17)_\text{sys}(11)_\text{eff}[27]_\text{total}.\notag
\end{align}
We observe that $a_e^\text{HLbL}$ becomes significantly more stable upon change of the matching variables than $a_\mu^\text{HLbL}$, especially for the longitudinal component this uncertainty is negligible. In the case of the electron, the low-energy contributions dominate to such an extent that the entire contribution from energy regions affected by SDCs and their implementation is small, resulting in less sensitivity to the details of the transition. To the contrary, for $a_\tau^\text{HLbL}$, the largest contribution arises from the pQCD and OPE regions, but the pseudoscalar poles and other low-energy contributions still play a relevant role, so that the relative matching  uncertainty even increases a little. Despite these largely expected changes in the relative importance of the various error components, the final uncertainty proves remarkably similar, we find a precision of $\{7,8,8\}\%$ for the light-quark contributions to $a_\ell^\text{HLbL}$,  $\ell\in\{e,\mu,\tau\}$.

\section{Charm loop}
\label{sec:charm}

\begin{figure}[t!]
     \centering
     \includegraphics[width=\linewidth]{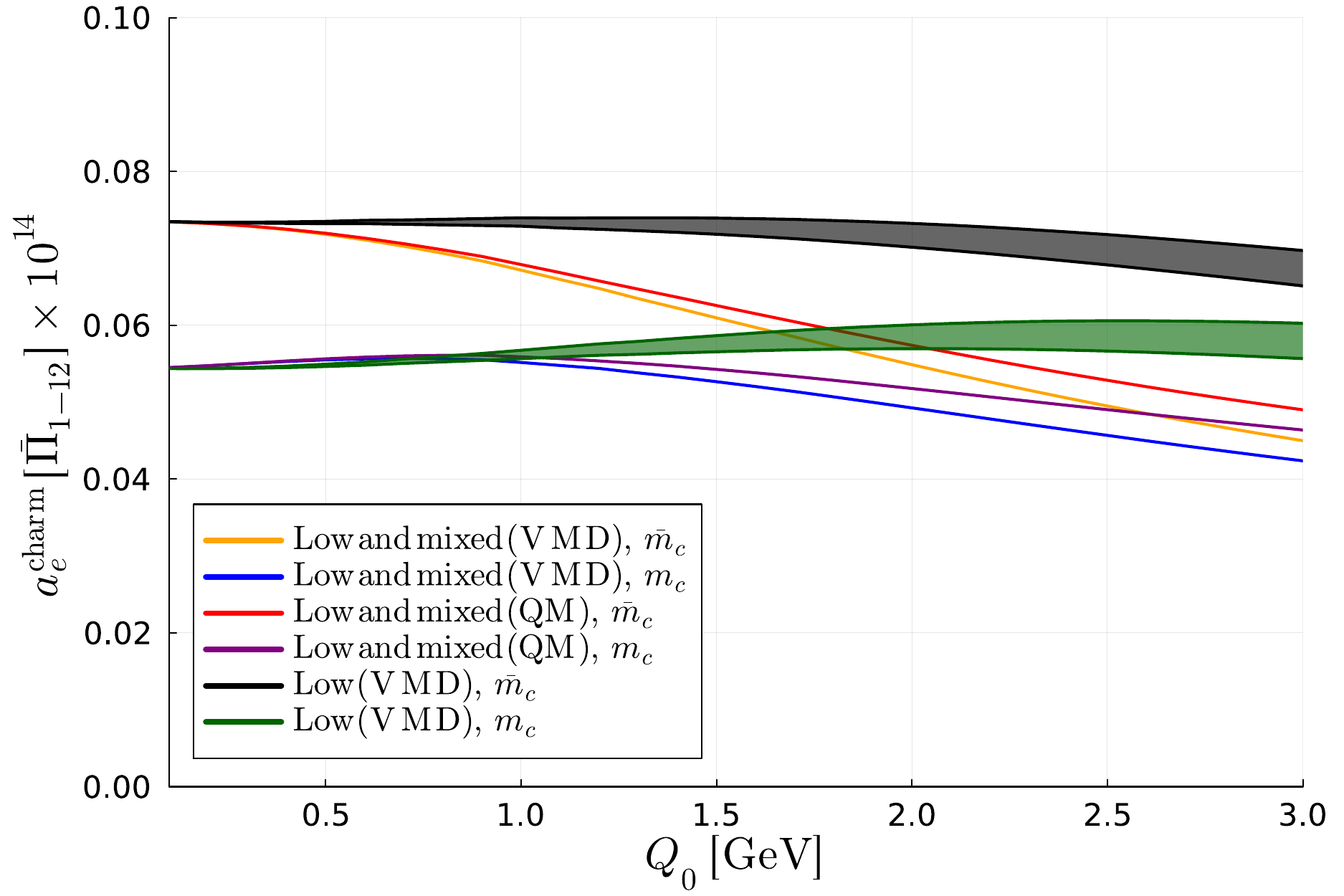}
     \includegraphics[width=\linewidth]{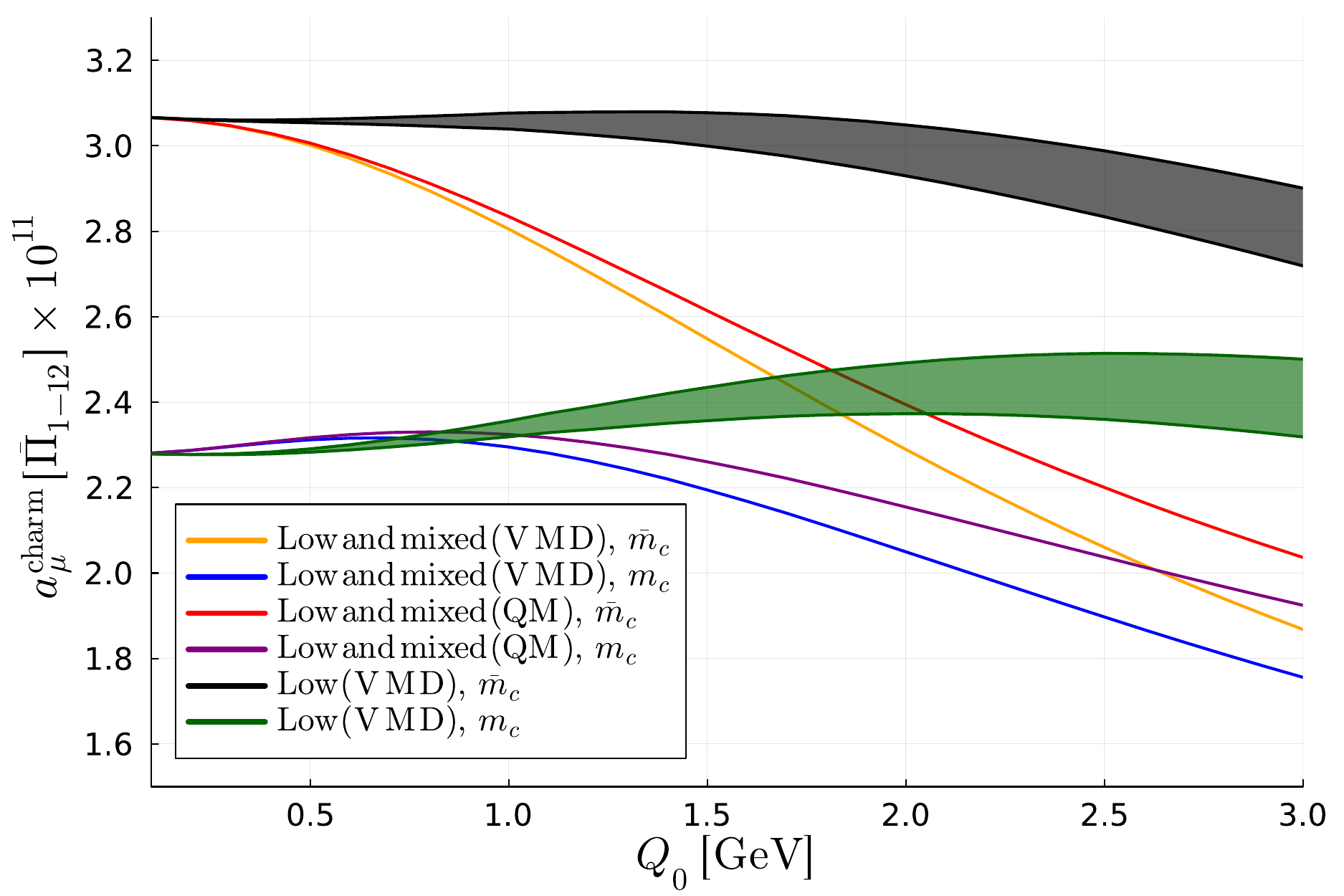}
     \includegraphics[width=\linewidth]{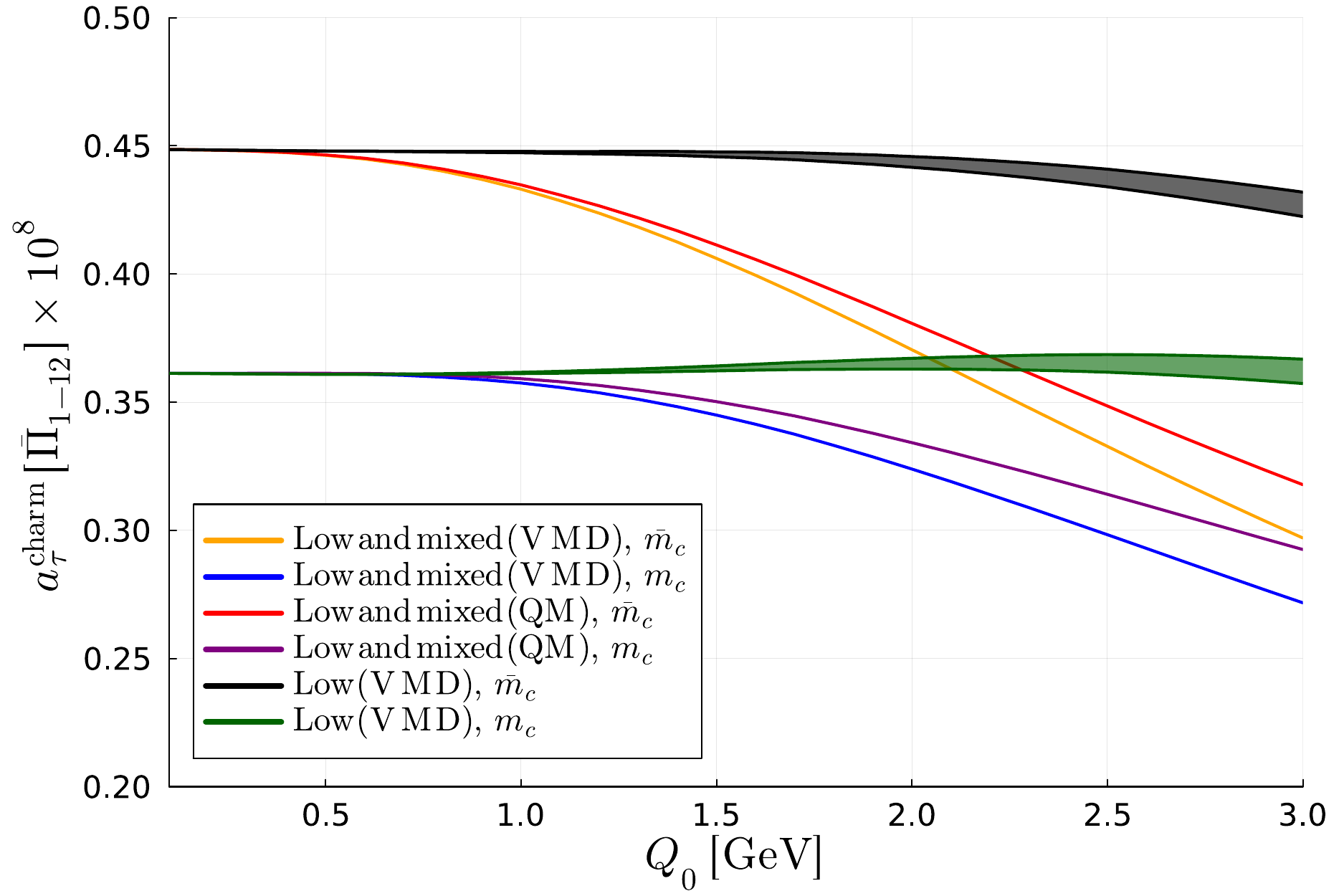}
     \caption{Charm-loop contributions to $a_e^\text{HLbL}$ (upper), $a_\mu^\text{HLbL}$ (middle), and $a_\tau^\text{HLbL}$ (lower), as a function of the scale $Q_0$ below which the hadronic description is used. The parameters for $\eta_c(1S)$, $\eta_c(2S)$ are taken from Ref.~\cite{ParticleDataGroup:2024cfk}, see also Ref.~\cite{Hoferichter:2025yih}. The curves refer to the scenarios (i)--(iii) defined in the main text, with $\overline{\text{MS}}$ charm mass $\bar m_c$ and pole mass $m_c$, respectively.}
     \label{fig:charm}
 \end{figure}

For the charm loop, the estimate $a_\mu^{\text{HLbL}}[c]=3(1)\times 10^{-11}$~\cite{Colangelo:2019uex} was based on the pQCD quark loop evaluated with $\overline{\text{MS}}$ mass $\bar m_c(m_c)\simeq 1.27\GeV$. In the case of the $VVA$ correlator, $\alpha_s$ corrections to the massive quark loop have been calculated~\cite{Melnikov:2006qb}, in a scheme that corresponds to the pole mass $m_c\simeq 1.48\GeV$. For HLbL scattering, however, the $\alpha_s$ corrections from Ref.~\cite{Bijnens:2021jqo} pertain to the massless case, so that the difference between $m_c$ and $\bar m_c$ constitutes a higher-order effect in $\alpha_s$. To estimate the uncertainty, the overall size of the $\eta_c$ pole was adopted in Ref.~\cite{Colangelo:2019uex}, without an attempt to match perturbative and hadronic descriptions. In particular, the resulting uncertainty safely covers the difference between $m_c$- and $\bar m_c$-based evaluations as well.

In view of the relevance of the charm loop for $a_\tau^\text{HLbL}$, we attempt a slightly improved evaluation here. To this end, we replace the pQCD quark loop below a scale $Q_0$ with the hadronic description based on $\eta_c(1S)$ and $\eta_2(2S)$, see Ref.~\cite{Hoferichter:2025yih},\footnote{We also considered the experimental uncertainties for the two-photon couplings of $\eta_c(1S)$ and $\eta_c(2S)$~\cite{ParticleDataGroup:2024cfk}. For the latter, uncertainties are rather sizable, but also for the former a tension with recent lattice-QCD calculations~\cite{Colquhoun:2023zbc,Meng:2021ecs} and the BESIII measurement~\cite{BESIII:2024rex} exists, suggesting an increased two-photon decay width.} for different scenarios, making the replacement (i) only in the low-energy region $Q_i\leq Q_0$, (ii) also in the mixed regions, (iii) as (ii) but with a quark-model TFF instead of a vector-meson-dominance (VMD) one; and all three for both mass schemes. The results in Fig.~\ref{fig:charm} show that scenario (i) is remarkably stable over a wide range of $Q_0$, suggesting that the pQCD quark loop matches well onto the hadronic description in the low-energy region, and rendering the mass scheme for $m_c$ the dominant source of uncertainty. For the mixed regions, we observe that the matching deteriorates rather quickly, partly due to the too-fast asymptotic decrease of the TFFs, as confirmed by the fact that (iii) behaves better than (ii). From the difference between the two mass schemes in Fig.~\ref{fig:charm}, we quote the following values for the charm-loop contributions
\begin{align}
\label{charm}
    a_e^\text{HLbL}[c]&=0.06(2)\times 10^{-14},\notag\\
    a_\mu^\text{HLbL}[c]&=2.7(8)\times 10^{-11},\notag\\
    a_\tau^\text{HLbL}[c]&=0.41(9)\times 10^{-8},
\end{align}
in agreement with the previous estimate $a_\mu^{\text{HLbL}}[c]=3(1)\times 10^{-11}$~\cite{Colangelo:2019uex}. The uncertainty around $(20\text{--}30)\%$ safely encompasses the $10\%$ shift due to $\alpha_s$ corrections found for the second-generation $VVA$ contribution in Ref.~\cite{Hoferichter:2025yih} (which also there roughly mirrors the difference in evaluations with $\overline{\text{MS}}$ and pole mass), suggesting that the resulting uncertainty in Eq.~\eqref{charm} due to $\alpha_s$ corrections may be an overestimate.

\section{Conclusions}
\label{sec:conclusions}

Combining the light-quark and charm contributions from Eqs.~\eqref{result_uds_e}, \eqref{result_uds_tau}  and~\eqref{charm}, we obtain
\begin{align}
    a_e^\text{HLbL}&=3.51(23)\times 10^{-14},\notag\\
    a_\tau^\text{HLbL}&=3.77(29)\times 10^{-8},
\end{align}
which constitutes our main result. Our evaluations are in agreement with Eqs.~\eqref{ae_HLbL_Jegerlehner} and~\eqref{atau_HLbL_Passera}, respectively, but considerably more precise while including a full error analysis. In particular, the uncertainty in $a_e^\text{HLbL}$ is pushed to a level two orders of magnitude below the current direct measurement~\eqref{ae_exp}. For the error propagation, we observe a slight  reduction for the electron, as the low-energy contributions, especially the pseudoscalar poles, dominate, so that the intricacies of matching to the asymptotic regions become less relevant. In contrast, for the $\tau$ lepton, the largest contribution arises from the pQCD and OPE regimes, but they do not dominate to the extent that the uncertainty from the transition region would be suppressed. Overall,  the relative precision therefore comes out almost identical for all three leptons, $\{7,8,8\}\%$ for $a_\ell^\text{HLbL}$,  $\ell\in\{e,\mu,\tau\}$.

\section*{Acknowledgments}

We thank Simon Holz for providing input for the $\eta$, $\eta'$ poles corresponding to the exact results of Refs.~\cite{Holz:2024lom,Holz:2024diw}.
Financial support by the SNSF (Project Nos.\ TMCG-2\_213690, PCEFP2\_194272, and PCEFP2\_181117) is gratefully acknowledged.

\bibliographystyle{apsrev4-1_mod}
\balance
\biboptions{sort&compress}
\bibliography{amu}

\end{document}